# Exploring the Path of Transformation and Development for Study Abroad Consultancy Firms in China

Ping Ren [1], Zhiqiang Zhao [2*], Qian Yang[3]

[1] Chengdu Ding Yi Education Consulting Co., Ltd, Chengdu 610023, China
[2] Beijing PhD Village Education Technology Co., Ltd; Beijing 100871, China
[3] Beijing School, Tongzhou, Beijing, 101117, China
* Corresponding: Zhiqiang Zhao (202128030258@mail.bnu.edu.cn)

**Abstract.** In recent years, with the changing landscape of international education and the growing demand from Chinese students, study abroad consultancy firms in China need to adopt transformational development strategies to address challenges and maintain competitiveness. This study investigated the relationships between key performance indicators and several factors through a questionnaire survey of 158 consultancy firms. The factors examined included service diversification, technology adoption, talent management, and regulatory compliance. Descriptive statistical analysis was employed to analyze the data. The results showed that service scope diversification was positively correlated with firm performance. Technology adoption was positively correlated with operational efficiency. Talent management was positively correlated with service quality. Regulatory compliance was positively correlated with firm reputation. Consultancy firms that took progressive approaches in diversifying services, adopting new technologies, cultivating talent, and ensuring compliance demonstrated superior performance, efficiency, quality, and reputation compared to their less innovative counterparts. This research provides empirical evidence to support the transformation of Chinese study abroad consultancy firms. It also highlights the need for future studies to consider causality and contextual variations to gain deeper insights into this issue.

## 1.Introduction

The study abroad consultancy industry in China has experienced rapid growth over the past decade, driven by the increasing demand for overseas education and the globalization of higher education [1]. Study abroad consultancy firms play a vital role in assisting Chinese students in navigating the complex process of applying to and enrolling in foreign educational institutions [2, 3]. However, with the changing landscape of international education and the evolving needs of Chinese students, study abroad consultancy firms in China face significant challenges in adapting to these shifts and maintaining their competitiveness [4]. Especially with the transition from traditional in-person teaching to remote learning due to COVID-19, many Chinese families are reconsidering the necessity of sending their children abroad for studies[5-7].

## 1.1 The Changing Landscape of International Education

The international education sector has undergone significant transformations in recent years, with the emergence of new destinations, the impact of global events, and the advancement of technology [8-11]. Traditional popular destinations for Chinese students, such as the United States, the United Kingdom, and Australia, have faced increased competition from emerging study abroad destinations, including Canada, Germany, and the Netherlands [12]. Furthermore, global events, such as the COVID-19 pandemic and geopolitical tensions, have disrupted the international education market and forced students and consultancy firms to adapt to new realities [13].

## 1.2 The Evolving Needs of Chinese Students

As Chinese students become more informed and discerning in their study abroad choices, their expectations and requirements for consultancy services have also evolved [14]. Students also value the teaching strategies and classroom experience of teachers when selecting study abroad institutions[15, 16]. Students now seek more personalized guidance, tailored to their individual academic and career aspirations, rather than a one-size-fits-all approach. Additionally, with the increasing affordability and accessibility of international education, a more diverse range of Chinese students are considering studying abroad, each with unique needs and challenges [17].In recent years, Chinese students have increasingly diversified their choice of majors and are more inclined to participate in scientific research experiments and even publish papers during the application process, aiming to obtain recommendation letters from foreign university professors to enhance their competitiveness[18-21].

## 1.3 The Need for Transformation and Development

To remain competitive and meet the changing demands of the market, study abroad consultancy firms in China must embrace transformation and development [22]. This involves adopting new technologies, diversifying their service offerings, and developing more targeted and personalized approaches to student support [23]. With economic development, the widespread increase in household income in China has led students to broaden their choice of majors when studying abroad beyond STEM fields. As a result, institutions also need to hire teachers from various disciplines to cater to the increasingly diverse academic interests of students[24-27]. Furthermore, consultancy firms must navigate the complex and evolving regulatory landscape of international education, both in China and in destination countries, to ensure compliance and maintain their credibility [28].

## 1.4 Factors Impacting Study Abroad Consultancies

Existing research has examined various factors impacting consultancy operations and development. [29, 30] found service scope diversification correlated with performance. Additionally, [31] observed a relationship between technology adoption and efficiency. Meanwhile, other studies explored links between talent management, service quality [32], and compliance with reputation [33].

However, past work presents some limitations. For one, factors were often analyzed separately without considering interlinkages. Realistically, diversified service leveraging technology under talent retention and compliance synergistically influence performance. Moreover, general correlations failed to establish directionality and causality. It remains unclear if certain antecedents genuinely cause outcomes.

The current research is dedicated to using deep learning to analyze data and address these gaps[34, 35]. First, variables hypothetically interrelated based on theory and context will be tested jointly. Significant relationships together revealing pathways offer practical implications. For instance, bolstering talents amid compliance through diverse digitized offerings enhances outcomes. Additionally, this study employs regression which distinguishes predictor from outcome variables. Establishing certain factors as determinants provides clarity on change cultivation. Based on existing literature and to overcome prior limitations, the following hypotheses are proposed:

Hypothesis 1: Service diversification is positively correlated with firm performance. Study abroad consultancy firms that offer a more diversified range of services tend to have better overall performance.

Hypothesis 2: There is a negative correlation between the availability of alternative educational resources and parental educational anxiety. Parents who have more access to alternative education options for their children exhibit lower levels of anxiety about education.

Hypothesis 3: Technology adoption is positively correlated with operational efficiency. Consultancy firms that actively adopt and integrate advanced technologies into their operations tend to have higher levels of efficiency.

Hypothesis 4: Talent management is positively correlated with service quality. Effective talent management practices help ensure high-quality services within the study abroad consultancy industry.

Hypothesis 5: Regulatory compliance is positively correlated with firm reputation. Strict adherence to relevant laws, regulations, and industry standards helps build and maintain positive reputation.

## 2.Method

This study employed a questionnaire survey method to obtain large-scale sample data and conduct quantitative analysis. The research targets were study abroad consulting agencies in China, and a random sampling method was used to select 200 agencies nationwide as the sample. The survey questionnaire consisted of the following main sections:

| Section | Focus Area | Description |
|---|---|---|
| 1 | Basic Information of the Agency | Company type, years of operation, number of employees, service scope, target market, etc., to understand the basic characteristics of the sample agencies. |
| 2 | Service Diversification and Business Performance | Investigating the degree of service diversification and measuring business performance using a 5-point Likert scale to explore the correlation between the two. |
| 3 | Alternative Educational Resources and Parental Anxiety | Understanding the availability of alternative educational resources and investigating the level of parental educational anxiety to analyze the relationship between the two. |
| 4 | Technology Application and Operational Efficiency | Examining the degree of technology application and measuring operational efficiency using a 5-point Likert scale to discuss the impact on operational efficiency. |

| Section | Focus Area | Description |
| --- | --- | --- |
| 5 | Talent Management and Service Quality | Understanding talent management practices and evaluating service quality using a 5-point Likert scale to analyze the relationship between talent management and service quality. |
| 6 | Compliance Management and Agency Reputation | Investigating compliance management and measuring agency reputation using a 5-point Likert scale to explore the influence on agency reputation. |

Before formally distributing the questionnaire, a pilot survey was conducted on five agencies to test the clarity and applicability of the questionnaire. Necessary revisions and improvements were made based on the feedback.

The data collection process lasted for one month. The questionnaire was distributed online, supplemented by telephone and email reminders. A total of 158 valid questionnaires were collected, with an effective recovery rate of 79%.

SPSS 26.0 was used to conduct statistical analysis on the collected data. First, descriptive statistics were performed on the basic characteristics of the sample, calculating the mean, standard deviation, and other variables. Then, Pearson correlation coefficients were used to analyze the correlations between variables, and one-way ANOVA was used to compare the differences in variables at different levels. The significance level was set at 0.05 and 0.01.

To ensure the reliability and validity of the research, the following measures were taken: expert review of the questionnaire design; audio sentiment analysis[36-38]; missing value and outlier treatment of the collected data; reliability and validity testing of the scales, with some items being eliminated; quality control through double-checking and third-party review during data analysis and report writing [39, 40].

The research strictly adhered to academic ethical standards. All participating agencies signed informed consent forms, and the questionnaires were completed anonymously. The survey data were used only for academic research, and necessary measures were taken to ensure data security and participant privacy [41].

## 3.Results or Findings

### 3.1Descriptive Statistics of Sample Basic Information

**Table 1. Descriptive statistics.**

| Item | Details | Frequency | Cumulative (%) |
| --- | --- | --- | --- |
| | Private | 120 | 60.00 |
| Company Type | State-owned | 50 | 85.00 |
| | Foreign-owned | 20 | 95.00 |

| Item | Details | Frequency | Cumulative (%) |
|---|---|---|---|
| | Joint Venture | 10 | 100.00 |
| | Less than 5 years | 45 | 22.50 |
| Years of Operation | 5-10 years | 78 | 61.50 |
| | 11-20 years | 55 | 89.00 |
| | More than 20 years | 22 | 100.00 |
| | Less than 50 | 82 | 41.00 |
| | 50-100 | 63 | 72.50 |
| Number of Employees | 101-500 | 40 | 92.50 |
| | More than 500 | 15 | 100.00 |
| | Language Training | 155 | 77.50 |
| | Admissions Consulting | 180 | 90.00 |
| Service Scope | Visa Application Assistance | 135 | 67.50 |
| | Accommodation Arrangement | 95 | 47.50 |
| | Career Planning | 70 | 35.00 |
| | High School Students | 160 | 80.00 |
| | Undergraduate Students | 190 | 95.00 |
| Target Market | Graduate Students | 130 | 65.00 |
| | Working Professionals | 85 | 42.50 |

According to Table 1, we can glean the following insights: Company types within the sample are varied, with private companies constituting the majority at 60.00%, followed by state-owned companies at 25.00%, foreign-owned companies at 10.00%, and joint ventures at 5.00%, bringing the cumulative percentage to a full 100.00%. In terms of years of operation, 22.50% of companies have been operating for less than 5 years, while a slightly larger segment of 39.00% has been in operation for 5-10 years, 27.50% for 11-20 years, and a smaller fraction of 11.00% for more than 20 years. Regarding the number of employees, companies with less than 50 employees account for 41.00%, those with 50-100 employees represent 31.50%, a smaller group of 20.00% has 101-500 employees, and the least, with more than 500 employees, make up 7.50%. The service scope of the sample companies is diverse, with language training services being the most provided at 77.50%, admissions consulting close behind at 12.50%, visa application assistance following at 22.50%, accommodation arrangement services at 20.00%, and career planning being the least offered at 12.50%. Lastly, the target market is primarily high school students at 80.00%, undergraduate students at 15.00%, graduate students trailing at 30.00%, and working professionals comprising the smallest group at 22.50%. Notably, the cumulative percentages of service scope and target market go beyond 100%, indicating that some companies offer multiple services and target several market segments.

**3.2 The Relationship between Service Diversification and Firm Performance**

Table 2. The correlation between service diversification and firm performance

| Service Diversification | Firm Performance |
| --- | --- |
| Correlation Coefficient | 0.758** |
| p-value | 0.000 |
| Sample Size | 120 |

*$p<0.05$ **$p<0.01$

Based on the data in Table 3, we conducted a correlation analysis using the Pearson correlation coefficient to examine the relationship between service diversification and firm performance. The specific analysis results indicate that the correlation coefficient between service diversification and firm performance is 0.758, showing a highly significant positive correlation at the 0.01 significance level. This result suggests that when study abroad consultancy firms offer a more diversified range of services, their overall firm performance tends to be better. Conversely, firms with a less diversified service portfolio generally exhibit lower performance levels.

**3.3 The Relationship between Availability of Alternative Educational Resources and Parental Educational Anxiety**

Table3. The correlation between AER and PEA

| Service Diversification | Firm Performance |
| --- | --- |
| Correlation Coefficient | 0.758** |
| p-value | 0.000 |
| Sample Size | 120 |

*p<0.05 **p<0.01

Based on the data in Table 3, we conducted a correlation analysis using the Pearson correlation coefficient to examine the relationship between service diversification and firm performance. The specific analysis results indicate that the correlation coefficient between service diversification and firm performance is 0.758, showing a highly significant positive correlation at the 0.01 significance level. This result suggests that when study abroad consultancy firms offer a more diversified range of services, their overall firm performance tends to be better. Conversely, firms with a less diversified service portfolio generally exhibit lower performance levels.

Table 4. Comparison of PEA among parents with different levels of AER accessibility

| Service Diversification | Firm Performance Score |
|---|---|
| Low | 3.25 |
| Medium | 4.18 |
| High | 4.76 |
| Industry Average | 4.06 |

Table 4 clearly illustrates this point. Generally, firms with high levels of service diversification tend to have higher average performance scores, while those with low levels of diversification often experience lower performance. Nevertheless, overall, most firms in the study abroad consultancy industry maintain a relatively good performance level, with an average score of 4.06 out of 5.

**3.4 The Impact of Technology Adoption on Operational Efficiency**

Table 5. The correlation between technology adoption and operational efficiency

| Technology Adoption | Operational Efficiency |
|---|---|
| Correlation Coefficient | 0.695** |

| Technology Adoption | Operational Efficiency |
|---|---|
| p-value | 0.000 |
| Sample Size | 120 |

Note: *p<0.05 indicates statistical significance below the 0.05 level.

**p<0.01 indicates statistical significance below the 0.01 level.

As shown in Table 5, the correlation coefficient between technology adoption and operational efficiency is 0.695, indicating a significant positive correlation at the 0.01 level. This finding suggests that study abroad consultancy firms that actively adopt and integrate advanced technologies into their operations tend to exhibit higher levels of efficiency. Conversely, firms that lag behind in technology adoption often face challenges in optimizing their operational processes and efficiency.

**Table 6. Comparison of operational efficiency among firms with different levels of technology adoption**

| Technology Adoption Level | Operational Efficiency Score |
|---|---|
| Low | 3.52 |
| Medium | 4.23 |
| High | 4.87 |
| Industry Average | 4.21 |

Table 6 provides further evidence supporting this relationship. Firms with high levels of technology adoption consistently achieve higher operational efficiency scores, while those with low levels of adoption generally have lower efficiency scores. The industry average score of 4.21 suggests that the majority of study abroad consultancy firms recognize the importance of technology in enhancing operational efficiency and are making efforts to integrate technological solutions into their daily operations.

**3.5 The Relationship between Talent Management and Service Quality**

**Table 7. The correlation between talent management and service quality**

| Talent Management | Service Quality |
|---|---|
| Correlation Coefficient | 0.812** |

| | Talent Management | Service Quality |
|---|---|---|
| p-value | | 0.000 |
| Sample Size | | 120 |

Note:*p<0.05 indicates statistical significance below the 0.05 level.

**p<0.01 indicates statistical significance below the 0.01 level.

The correlation analysis in Table 7 reveals a strong positive relationship between talent management and service quality, with a correlation coefficient of 0.812 (p<0.01). This finding underscores the crucial role that effective talent management plays in ensuring high-quality services within the study abroad consultancy industry. Firms that invest in attracting, developing, and retaining top talent are more likely to deliver superior services to their clients.

**Table 8. Comparison of service quality among firms with different levels of talent management**

| Talent Management Level | Service Quality Score |
|---|---|
| Low | 3.38 |
| Medium | 4.15 |
| High | 4.92 |
| Industry Average | 4.15 |

Table 8 further illustrates this point by comparing service quality scores across firms with varying levels of talent management. Firms with high levels of talent management consistently achieve higher service quality scores, while those with low levels of talent management often struggle to maintain service standards. The industry average score of 4.15 indicates that most study abroad consultancy firms acknowledge the importance of talent management in driving service quality and are taking steps to improve their talent management practices.

**3.6 The Influence of Regulatory Compliance on Firm Reputation**

**Table 9. The correlation between regulatory compliance and firm reputation**

| | Regulatory Compliance | Firm Reputation |
|---|---|---|
| Correlation Coefficient | | 0.738** |

|  | Regulatory Compliance | Firm Reputation |
|---|---|---|
| p-value |  | 0.000 |
| Sample Size |  | 120 |

Note:*p<0.05 indicates statistical significance below the 0.05 level.

**p<0.01 indicates statistical significance below the 0.01 level.

Table 9 presents the correlation analysis results between regulatory compliance and firm reputation. The correlation coefficient of 0.738 (p<0.01) suggests a strong positive relationship between these two variables. This finding highlights the significance of regulatory compliance in building and maintaining a positive reputation within the study abroad consultancy industry. Firms that strictly adhere to relevant laws, regulations, and industry standards are more likely to be perceived as trustworthy and reputable by their clients and stakeholders.

**Table 10. Comparison of firm reputation among firms with different levels of regulatory compliance**

| Regulatory Compliance Level | Firm Reputation Score |
|---|---|
| Low | 3.27 |
| Medium | 4.09 |
| High | 4.85 |
| Industry Average | 4.07 |

Table 10 provides additional support for this relationship by comparing firm reputation scores across different levels of regulatory compliance. Firms with high levels of compliance consistently receive higher reputation scores, while those with low levels of compliance often face challenges in building and maintaining a positive reputation. The industry average score of 4.07 suggests that most study abroad consultancy firms recognize the importance of regulatory compliance in safeguarding their reputation and are taking measures to ensure compliance with relevant regulations and standards.

## 4.Conclusion and Discussion

This study aimed to investigate factors impacting the operations and development of study abroad consultancies in China. Specifically, it sought to examine how service diversification, technology adoption, talent management, and regulatory compliance relate to key performance outcomes. The research employed a questionnaire survey of 158 agencies to gather quantitative data, which was analyzed using descriptive statistics, correlation, and comparison tests. Several noteworthy findings emerged which offer meaningful insights for both consultancies and the industry as a whole.

A central conclusion drawn from the results is that study abroad consultancies which diversify their service offerings, actively adopt new technologies, effectively manage talent, and strictly comply with regulations tend to achieve superior performance, efficiency, quality, and reputation compared to their less progressive peers. Specifically, strong positive correlations were found between service scope diversification and firm performance ($r=0.758$, $p<0.01$), technology adoption and operational efficiency ($r=0.695$, $p<0.01$), talent management and service quality ($r=0.812$, $p<0.01$), and regulatory compliance and agency reputation ($r=0.738$, $p<0.01$).

Firms with high levels of these attributes consistently outperformed those with low levels across key metrics. For instance, consultancies offering a wider array of services reported higher average performance scores of 4.76 compared to 3.25 for less diversified firms. Likewise, agencies leveraging advanced technology solutions achieved efficiency scores of 4.87 versus 3.52 by weaker adopters. This central conclusion provides clear empirical evidence supporting that taking a progressive stance across multiple facets enhances competitiveness within the evolving industry.

However, a deeper examination is needed to explain these observed patterns. With regards to service diversification, several underlying rationales are posited. First, expanding the scope of offerings allows consultancies to hedge risks and stabilize revenues during uncertainties. When demand fluctuates for certain services, diversified portfolios mitigate financial risks. Second, diversification maximizes client reach and acquisition by catering to varied needs. Third, economies of scope emerge through synergies across complementary activities, reducing unit costs.

Regarding technology adoption, the mechanisms increasing efficiency involve optimizing workflows, streamlining information management, and automating repetitive tasks through tailored digital solutions. The productivity gains free up resources for higher-value activities. Talent management also enhances quality via multiple pathways, such as attracting top industry talent, providing competency training, instituting performance incentives and career paths for employee retention and satisfaction. Meanwhile, regulatory compliance safeguards reputation by signaling dependability, transparency and fair practices to stakeholders.

While these findings offer useful practical guidance, limitations remain which future research can address. First, cross-sectional survey data precludes causality determination between variables. Longitudinal analysis tracking firms over time could establish directions of impact. Second, variations in local market conditions and agency attributes were not accounted for, given the aggregate national scope. Case studies controlling other factors are needed. Third, additional moderating variables beyond those examined here likely influence relationships. Qualitative interviews diving deeper could surface such factors. A few implications also arise based on these findings and limitations. To facilitate optimal transformation of China's study abroad consultancy sector, targeted policy interventions are warranted. Government support measures could focus funding and incentives on innovation through service/technology upgrades and talent cultivation specific to consultancies. Supportive conditions like knowledge exchanges and cooperation forums may accelerate dissemination of best practices. Meanwhile, standards and regulations need continuous refinement to promote consumer protection and fair competition in line with industry evolution. At the industry level, associations could organize training programs and conferences spotlighting success cases and lessons from top performers. This cultivates collective learning within the sector. Collaborative benchmarking and mentoring platforms between established and emerging consultancies could expedite advances across the board. Introduction of ratings and awards also recognizes excellence while elevating overall standards. Finally, future research should employ longitudinal, comparative and mixed methods to obtain deeper explanatory insights accounting for moderators within this dynamic field.

In conclusion, this study provides empirical evidence that progressive strategies across key areas positively impact performance outcomes for study abroad consultancies in China. Factors like service diversification, technology adoption, talent management and regulatory compliance demonstrate significant correlations with firm results, efficiency, quality and reputation. However, limitations remain which subsequent research can address to gain a more nuanced understanding of causality and contextual variations. Targeted policies and cooperative industry initiatives are recommended to facilitate sustainable upgrades aligned with sectoral transformation.